\documentclass[amsmath, amssymb, onecolumn]{revtex4-2}
\usepackage{etoolbox}
\AtBeginEnvironment{thebibliography}{\hypersetup{hidelinks}}
\usepackage{graphicx}
\usepackage{subcaption}
\usepackage{dcolumn}
\usepackage{bm}

\usepackage[utf8]{inputenc}
\usepackage[T1]{fontenc}
\usepackage{mathptmx}
\usepackage{etoolbox}
\usepackage{lipsum}
\usepackage[colorlinks=true, linkcolor=black, citecolor=black, urlcolor=blue]{hyperref}

\newcommand{\eps}{\varepsilon}
\newcommand{\beps}{\epsilon}
\newcommand{\cres}{c^\text{res}}
\newcommand{\be}{\begin{equation}}
\newcommand{\ee}{\end{equation}}
\newcommand{\bea}{\begin{eqnarray}}
\newcommand{\eea}{\end{eqnarray}}
\newcommand{\tc}{(\beta \eps_\text{pp})_\text{c}}
\newcommand{\bepp}{\beta \eps_\text{pp}}
\newcommand{\xo}{X_\text{o}}
\newcommand{\xu}{X_\text{u}}
\newcommand{\mo}{m_\text{o}}
\newcommand{\muo}{m_\text{u}}

\makeatletter
\def\@email#1#2{%
 \endgroup
 \patchcmd{\titleblock@produce}
  {\frontmatter@RRAPformat}
  {\frontmatter@RRAPformat{\produce@RRAP{*#1\href{mailto:#2}{#2}}}\frontmatter@RRAPformat}
  {}{}
}%
\makeatother
\begin{document}


\title{(Dis-)appearance of liquid-liquid phase transitions in a heterogeneous activated patchy particle model and experiment}

\author{Furio Surfaro}
\affiliation{Institute of Applied Physics, University of T{\"u}bingen, 72076 T{\"u}bingen, Germany}
\author{Peixuan Liang}
\affiliation{Institute of Applied Physics, University of T{\"u}bingen, 72076 T{\"u}bingen, Germany}
\author{Hadra Banks}
\affiliation{Institute of Applied Physics, University of T{\"u}bingen, 72076 T{\"u}bingen, Germany}
\author{Fajun Zhang}
\affiliation{Institute of Applied Physics, University of T{\"u}bingen, 72076 T{\"u}bingen, Germany}
\author{Frank Schreiber*}
\affiliation{Institute of Applied Physics, University of T{\"u}bingen, 72076 T{\"u}bingen, Germany}
\email{frank.schreiber@uni-tuebingen.de}
\author{Martin Oettel*\thanks{Corresponding Author}}
\affiliation{Institute of Applied Physics, University of T{\"u}bingen, 72076 T{\"u}bingen, Germany}
\email{martin.oettel@uni-tuebingen.de}

\begin{abstract}
The ion-activated patchy particle model is an important theoretical framework to investigate
the phase behaviour of globular proteins in the presence of multivalent ions. In this work,
we study and highlight
the influence of patch heterogeneity on the extension, appearance and
disappearance of the liquid-liquid coexistence region of the phase diagram.
We demonstrate that within this model
the binding energy between salt ions and patches of different type is a key factor
in determining the phase behavior.
Specifically, we show under which conditions liquid-liquid phase separation (LLPS)
in these systems can appear or disappear for varying binding energy  and
ion-mediated attraction energy between ion-occupied and unoccupied patches. In particular we address the influence of the patch type dependence of these energies on the (dis)appearance of LLPS.
These results rationalize our new results on ion-dependent liquid-liquid phase separation in solutions of
bovine serum albumine with trivalent cations. 
In comparison with models with non-activated patches, where the gas-liquid transition disappears when the number of patches approaches two, we find the complementary mechanism that ions may shift the attractions from stronger to weaker patches (with an accompanying disappearance of the transition), if their binding energy to the patches changes.   
The results have implications for the understanding of charge-driven LLPS
in biological systems and its suppression. 
\end{abstract}

\maketitle

\section{Introduction}

Suspensions of structured colloidal particles\cite{kim2025patchy_stenciling,chen2011supracolloidal,shah2015actuation,choueiri2016surface,gong2017patchy,wang2012colloids}, globular proteins\cite{fusco2013crystallization,bergman2025dynamical,bianchi2011patchy,gnan2019patchy,doye2007controlling,audus2018valence}, and related systems are often
effectively modeled as patchy particles \cite{tavares2010equilibrium, bianchi2011patchy, bianchi2017limiting, pawar2010fabrication, staneva2015role, notarmuzi2025simulating}. These systems exhibit a rich phase behavior due to
their directional interactions and have been studied extensively, both for their theoretical interest
and practical relevance. In particular, patchy particle models, in which their
charged or chemically distinct surface regions act as interaction patches  \cite{roosen2014ion,de2011phase, sciortino2005glassy,russo2009reversible,russo2022physics,Fujihara2014,Jungwirth2012,Kumar2021,Kundu2013},
have proven useful in describing protein phase behavior, including
crystallization and phase separation, in particular liquid-liquid phase separation (LLPS),
the latter being intensely debated as a mechanism for molecular organization biological cells
\cite{brangwynne2013phase,hyman2014liquid,zhang2020liquid}.
%

{Experimentally, LLPS in solutions of globular proteins in the presence of trivalent salts (proteins-Me$^{3+}$-systems) has been observed for a range of systems.\cite{matsarskaia2020multivalent} The binodal of the LLPS can be determined by partitioning of both salt and protein into two coexisting phases,\cite{Zhang2012a,Zhang2014,Wolf2014,Zhang2017,Braun2017,Braun2018,matsarskaia2016cation,Maier2021a,surfaro2023alternative}, and the so determined coexistence region is bounded by a closed-loop in the salt-protein concentration plane.\cite{Zhang2012a, Wolf2014, matsarskaia2016cation} 
The employed proteins were bovine serum albumine (BSA), human serum albumine (HSA), beta-lactoglobuline (BLG), ovalbumin (OVA)  and Me$^{3+}$ ions included Y$^{3+}$, La$^{3+}$, Ho$^{3+}$, Gd$^{3+}$, Ce$^{3+}$, Yb$^{3+}$  with strong differences in the degree of LLPS between different combinations of proteins and ions. E.g., visual inspection and turbidity measurements show that in BSA solutions with HoCl$_3$ and GdCl$_3$, there is strong phase separation while with LaCl$_3$ a phase transition is absent at room temperature.\cite{Matsarskaia_2018_PhysChemChemPhys} 
The effective protein-protein interactions characterized using SAXS indicate a dominant short-ranged attraction, that accounts for the metastability of LLPS with respect to crystallization. \cite{ Zhang2012a, Wolf2014, Braun2017,Braun2018}
Studies of protein crystallization in the presence of multivalent metal ions suggest that the metal ions are not only used to induce crystallization, but are an integral part of the crystal lattice\cite{Zhang2011}. Structural analysis demonstrates specific binding of metal ions to surface-exposed glutamate and aspartate side chains contributed by different protein molecules in the crystal lattice\cite{Zhang2011}.  By bridging molecules in this manner, contacts between molecules are formed that enable the formation of a stable crystal lattice. Based on this bridging effect, an ion-activated patchy particle model was proposed to rationalize the phase behavior observed experimentally\cite{roosen2014ion}. 
Within this analytical approach proteins are modeled as particles with a finite number of patches per particle. Multivalent cations are modeled as bridging or linker particles which can bind to these patches and thereby activate them. Without cation-mediated protein-protein bridging, the interaction between proteins is modelled as hard-sphere repulsion. However, when an occupied patch interacts with an unoccupied one, an ion bridge forms and links the participating proteins with an attractive square-well attraction. 

Importantly, previous investigations \cite{roosen2014ion, surfaro2024ion} of the ion-activated  patchy particle model assumed particles with {\it equivalent} patches — an assumption that oversimplifies the inherently heterogeneous nature of many biological and synthetic systems.
%
Here, we present an extension of the model in an analytically tractable manner that aims to elucidate the effects of patch heterogeneity on the phase diagram. 
We generalize the
interaction energy by taking into account different kinds of patches with different binding probabilities,
which is a more realistic model for proteins with their inherent inhomogeneities and low symmetry.

Already for the case of two patch types, we find that the extension and location of
the binodal region can be tuned mainly by the binding specificity of the binders (salt ions) with the patches on the surfaces of the particles. When ions bind differently to the different patch types, attractions between occupied and unoccupied patches may shift between dominant and weak combinations of attractive patches, and the binodal region consequently changes size and location (and may also disappear completely). 
Experimentally, we demonstrate such a change in the binodal region by measurements on  BSA systems with YCl$_3$ vs. HoCl$_3$.

This paper is organized as follows.
In Sec.~\ref{sec:theory} we explain the theoretical background and how our specific model and its extensions are set up,
including the connection to Wertheim theory, which provides a theoretical framework
for predicting the thermodynamic properties of patchy particle systems.
Sec.~\ref{sec:exp} briefly describes the experimental procedure for obtaining phase diagrams in BSA solutions with YCl$_3$ vs. HoCl$_3$.  
Sec.~\ref{sec:results} is devoted to the results of our model with a detailed analysis of the appearance / disappearance
of the LLPS as a function of the model parameters and presents and rationalizes the experimental phase diagrams.
We conclude by a summary and a discussion of the implications for the phase behavior of protein-salt systems
and specifically the existence of LLPS in biology, which is presently a subject of intense debate \cite{brangwynne2013phase,hyman2014liquid,zhang2020liquid}

\section{Theory and model setup}
\label{sec:theory}

We treat proteins as spherical colloidal particles (hard spheres) which possess specific binding sites for trivalent ions (see Fig.~\ref{fig:sketch_model} for an illustration). The number of these binding sites (patches) is $M$. Since the protein surface is heterogeneous, patches can have different binding properties and will be distinguished by greek indices $\alpha, \beta...$. The number of patches of type $\alpha$ is given by $m_\alpha$. The trivalent ions (labelled by a latin index $i$) are assumed to generate an effective attraction between the proteins if they bind with patches of two different proteins, e.g. they act as linker particles between the colloidal particles.

A typical patchy particle model setup consists of a mixture of a colloidal species with $M$ patches (of possibly different type) with a linker species having 2 patches which is able to bind two colloidal particles. The behavior of such a system is mainly governed by a patch-ion  depending on ion species $i$ and patch type $\alpha$.    
Such particle-linker models have been investigated in the literature \cite{braz2021phase,gouveia2022linkers,howard2019structure}, with regard to phase behavior we note that e.g. Ref.~\cite{braz2021phase} finds liquid-liquid binodals in the form of closed loops in the particle density--linker density plane which is similar to the closed loops in the BSA-Me$^{3+}$ system. 

\begin{figure}[htbp]

\includegraphics[height=0.22\linewidth,keepaspectratio]{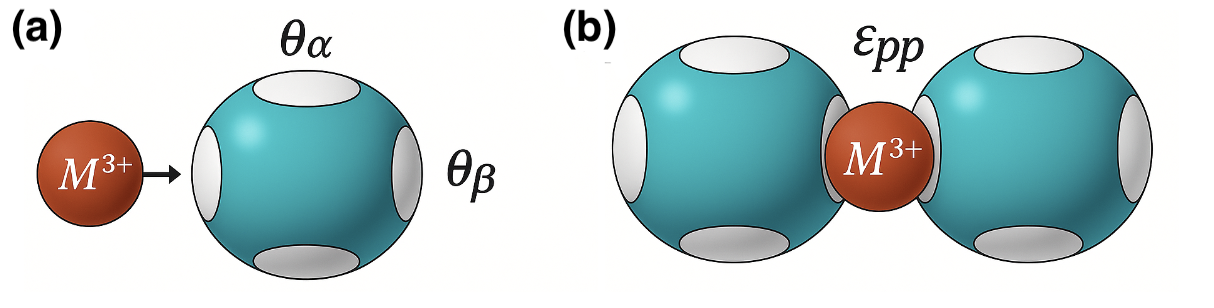}
\caption{(a) Metal ions are bound to patches of type $\alpha$, $\beta$ ... with probability $\theta_{\alpha}, \theta_{\beta}$.... (b) Occupied patches (of any type $\alpha$) interact attractively with unoccupied patches (of any type $\beta$). Averaging (see text) gives an effective patch-patch interaction energy $\eps_\text{pp}$ between occupied and unoccupied patches.}

  \label{fig:sketch_model}
\end{figure}
 
We employ a further approximation to the particle-linker picture by treating the average ion-binding occupancy of a single patch by the statistics of a two-
state system in the grand canonical ensemble, as introduced in Ref.~\cite{roosen2014ion}. The relative occupation $\theta^i_\alpha$ of a patch $\alpha$ by an ion $i$ is given by
by a Fermi-like distribution:
\begin{equation} \label{eq:binding}
    \theta^i_\alpha = \frac{1}{1+\exp\big(\beta({\beps_b^{i,\alpha} -\mu_{s,i})\big)}},
\end{equation}
where $\beta=1/(k_B T)$ is the inverse temperature, $\beps_b^{i,\alpha}$ is the binding energy between a salt ion of type $i$ and a patch $\alpha$. The thermodynamic control variable for the occupation is the ion chemical potential $\mu_{s,i}$ which for dilute systems can be approximated with that of an ideal gas with a reservoir concentration $\cres_i=c_0\exp(\beta \mu_{s,i})$ where $c_0$ is a reference concentration. The effective binding of two particles with the ion as a linker occurs through an attractive binding energy only between an ion-occupied (o) patch and an unoccupied (u) patch, $\eps_\text{uo}^{\alpha\beta}<0$, which depends on the type of patches involved. 
It appears that the elimination of ions as separate linker species has resulted in a multi-component model of particles with $0...M$ occupied patches (of possibly different types) which would complicate the determination of phase diagrams. However, following Ref.~\cite{roosen2014ion} we employ the mapping to an effective model with only one type of patch with the effective patch-patch (pp) interaction energy
\begin{align}
\beta\epsilon_{\mathrm{pp}}
&=
\frac{1}{M^{2}}
\sum_{\alpha,\beta}
m_\alpha m_\beta
\Big[
\beta\epsilon_\text{uu}^{\alpha\beta}\,(1-\theta^i_\alpha)(1-\theta^i_\beta)
+\beta\epsilon_\text{oo}^{\alpha\beta}\,\theta^i_\alpha\theta^i_\beta
\Big]
\notag\\
&\quad+
\frac{2}{M^{2}}
\sum_{\alpha,\beta}
m_\alpha m_\beta\,\beta\epsilon_\text{uo}^{\alpha\beta}\,\theta^i_\alpha(1-\theta^i_\beta).
\label{eq:beta_epp_general_symmetric}
\end{align}
where we set $\beta\epsilon^{\alpha\beta}_\text{uu}=\beta\epsilon^{\alpha\beta}_\text{oo}=0$, i.e. interactions between pairs of (un)occupied patches do not contribute to the attractions.
For the attractions, we assume independent association of occupied patches (fraction $\theta^i_\alpha$) with unoccupied patches (fraction  $1-\theta^i_\beta$), weighted with the fractions $m_\alpha/M$ and $m_\beta/M$ of patches $\alpha$ and $\beta$ on the particle surface. With this, the central elements of the generalized ion-activated patchy particle model are introduced; a schematic sketch is shown in Fig.~\ref{fig:sketch_model}.  
In App.~\ref{app:twobindingsites} we investigate an extended, effective model which does not rely on an averaged, effective patch-patch energy. In our context, it is numerically more demanding but gives a similar phenomenology. More generally, we have noted that the extended model is equivalent to certain linker-receptor models which describe funcionalized nanoparticle interactions with surfaces \cite{angioletti2020}.

The thermodynamics of this effective model with one type of patch is treated with Wertheim theory \cite{Wertheim1984a,Wertheim1984b,Wertheim1986} and is briefly summarized in App.~\ref{app:wertheim}. Wertheim theory allows us to calculate the binodal (see Fig.~\ref{fig:binodal} in App.~\ref{app:wertheim}) and gives a critical temperature (critical interaction strength) which we write in dimensionless form as $\tc<0$. If $\bepp<\tc$, the model shows phase separation: the effective patch-patch attraction is stronger than the one needed at the critical point and the system shows liquid-liquid phase separation. For given interaction parameters $\eps_\text{uo}^{\alpha\beta}$, $\eps_\text{pp}$ is a function of the occupancies $\theta^i_\alpha$ and has a minimum $\eps_\text{pp,min}$. For $\beta \eps_\text{pp,min} \le \beta  \eps_\text{pp} \le \tc$, the coexisting colloid (protein) packing fractions $\eta_1$ and $\eta_2$ are obtained from the Wertheim binodal.  
Eqs.~(\ref{eq:binding}) and (\ref{eq:beta_epp_general_symmetric}) can be solved for the ion chemical potential $\mu^i_s$ (or equivalently the salt reservoir concentration $\cres_i$), $\eta_1$ and $\eta_2$  thus depend on $\cres_i$ and the curves $\eta_1(\cres_i)$ and $\eta_2(\cres_i)$ define a closed loop for the binodal in the $\eta$-$\cres$ plane. To convert to physical salt concentrations $c_s$, it is assumed that the ideal ions and the colloids form a mixture of Asakura--Oosawa type \cite{roosen2014ion}:
\be
 c_s = \sum_\alpha m_\alpha \theta^i_\alpha \rho + \cres(\mu^i_s) (1 - \eta (1 + R_s/R)^3)
 \label{eq:cs}
\ee
Here, $R$ and $R_s$ denote the radius of colloids and ions, respectively, and the colloid density $\rho$ and packing fraction $\eta$ are connected by $\eta=\rho (4\pi R^3/3)$. The first term in Eq.~(\ref{eq:cs}) describes the ions bound to the protein surface patches and the second term gives the concentration of mobile ions which is smaller than $\cres$ since the free volume for ions is reduced due to the presence of the colloidal particles (proteins). This free volume is given up to linear order in $\eta$. It is seen that $c_s(\eta)$ is a linear function of $\cres$, thus the closed loop for the binodal in the $\eta$-$c_s$ plane will be tilted and stretched compared to the loop in the $\eta$-$\cres$ plane.

\section{Experiment}
\label{sec:exp}

BSA (product No. A7906), HoCl$_3$ and YCl$_3$,  were purchased from Sigma Aldrich and used as received.  For the stock solutions, protein and salt were dissolved in degassed Milli-Q H$_2$O (18.2$\,$M$\rm{\Omega}$cm conductivity). No buffer was used to avoid the effects of other co-ions. BSA has a molecular mass of 66.5\,kDa and an isoelectric point of $\rm{pI}=4.6$. All experiments were performed at room temperature ($23\pm 2\,^\circ$C).
Concentrations of protein stock solutions were determined by measuring the absorbance at 280 nm using a Cary 50 UV-vis spectrophotometer (Varian Inc.) with the software Cary WinUV.  The extinction coefficient of BSA is $E_{280}=0.667\,\rm{ml}/\left(\rm{mg\cdot cm}\right)$~\cite{sober1970handbook}.
For LLPS, a series of sample solutions with a fixed protein concentration of $c_p=120$ mg/ml (corresponding to a protein volume fraction $\eta_\mathrm{exp}^\text{ini} \approx 0.08$ by using $\eta_{\mathrm{exp}} = {c_p v_p}/{(1+c_p v_p)}$ and the specific volume $\nu_p=0.725$ ml/g) and varying salt concentrations $c_s^\text{ini}$ was prepared.  After phase separation, the volumes of the dense and dilute phases were recorded. The protein concentration in the dilute phase was determined using UV-visible spectroscopy, whereas the protein concentration in the dense phase was calculated using the lever rule.

\section{Results}
\label{sec:results}

Secs.~\ref{sec:single} and \ref{sec:two} discuss the generic behavior in the ion-activated patchy particle model with one and two types of patches, respectively. In Sec.~\ref{sec:exploop} the experimental phase diagram is presented along with a rationalization based on the model with  two types of patches.

\subsection{Single type of patch}
\label{sec:single}

\begin{figure}[ht!]
  \begin{center}
    \includegraphics[width=17cm]{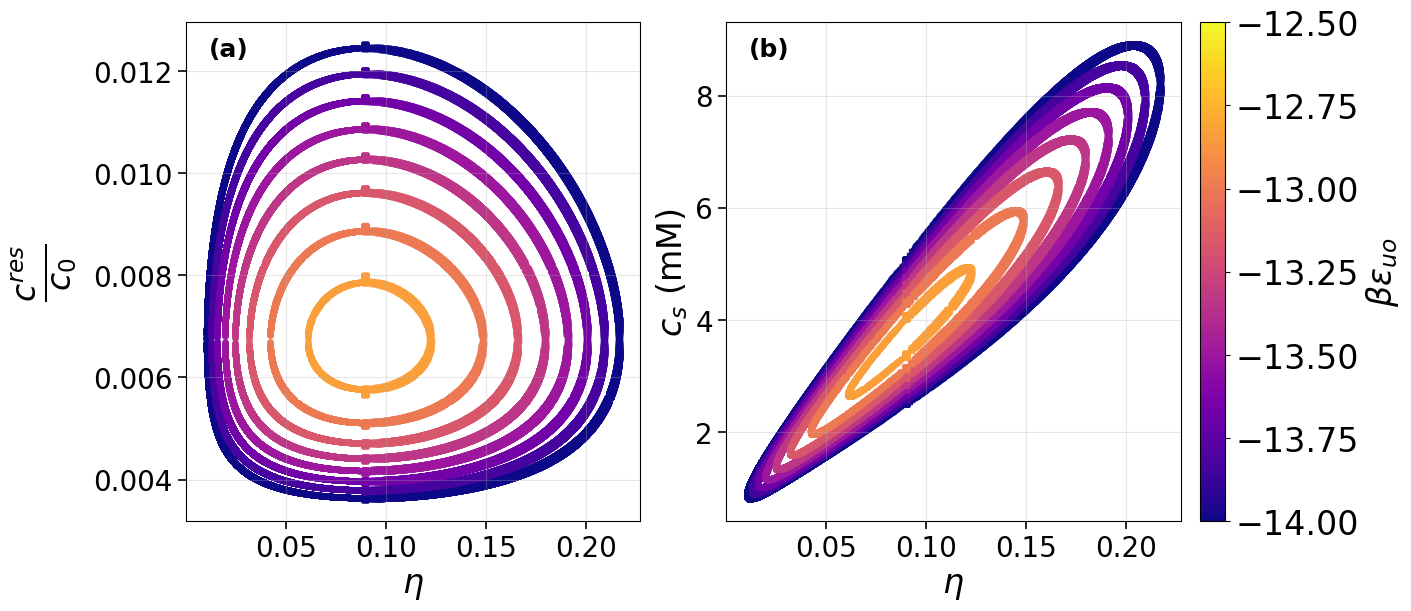}
    \end{center}
\caption{Change of binodal loop in the single-type-of-patch model upon variation of $\beta \eps_\text{uo}$ (``onion-shell behavior''). $\beta\eps_b= -5$, $c_0$ = 1M, $R =2.8$ nm, $R_s=R/18$. (a) $\eta$-$\cres$ plane (reservoir salt concentration). (b) $\eta$-$c_s$ plane (physical salt concentration). }
    \label{fig:onion}
\end{figure}

It is instructive to study briefly the shape of binodals and their parameter dependence in the simplest case of only a single type of patch. For this case, a particular choice of reasonable model parameters has been defined in Ref.~\cite{roosen2014ion} (see also App.~\ref{app:wertheim}), and it has been demonstrated that the binodal loop in the $\eta$-$c_s$ plane looks very similar to the experimental loop in a HSA-YCl$_3$ system at room temperature.

For the single type of patch (and one type of salt), we can drop all related indices, and  Eqs.~(\ref{eq:binding}) and (\ref{eq:beta_epp_general_symmetric}) become
\be
   \theta = \frac{1}{1+\exp\big(\beta({\beps_b -\mu_{s})\big)}}, \quad
   \eps_\text{pp} = 2\eps_\text{uo}\, \theta (1- \theta) \;,
  \label{eq:1patch}
\ee
and the phase diagram is essentially a function of $\beta \eps_\text{uo}$ and $\beta \beps_b$. The effective patch-patch energy becomes minimal (most attractive) for $\theta=1/2$.

In Fig.~\ref{fig:onion}, the behavior of the binodal loop upon variation of   $\beta \eps_\text{uo}$ is shown both in the $\eta$-$\cres$ and in the $\eta$-$c_s$ plane. Upon lowering $\beta \eps_\text{uo}$ (making the attraction between occupied and unoccupied patches stronger), the binodal widens in both directions in an onion-shell manner, which is easily understood from the Wertheim solution for the two coexisting packing fractions $\eta_1$ and $\eta_2$ and Eq.~(\ref{eq:1patch}). With stronger attraction, the maximal coexistence gap $\Delta \eta_\text{max}=\text{max}(\eta_2-\eta_1)$ widens, as does the gap between the two critical occupations $\theta_{c,i}$, which are obtained as the solutions of $\tc=2\beta \eps_\text{uo}\, \theta (1- \theta)  $. This result also results in a widening gap between the critical reservoir salt concentrations. This onion-shell behavior of the binodal loop is very similar to the experimentally observed behavior of the binodal upon variation of the temperature \cite{matsarskaia2016cation,Matsarskaia_2018_PhysChemChemPhys}. 

\begin{figure}[ht!]
    \begin{center}
    \includegraphics[width=16cm]{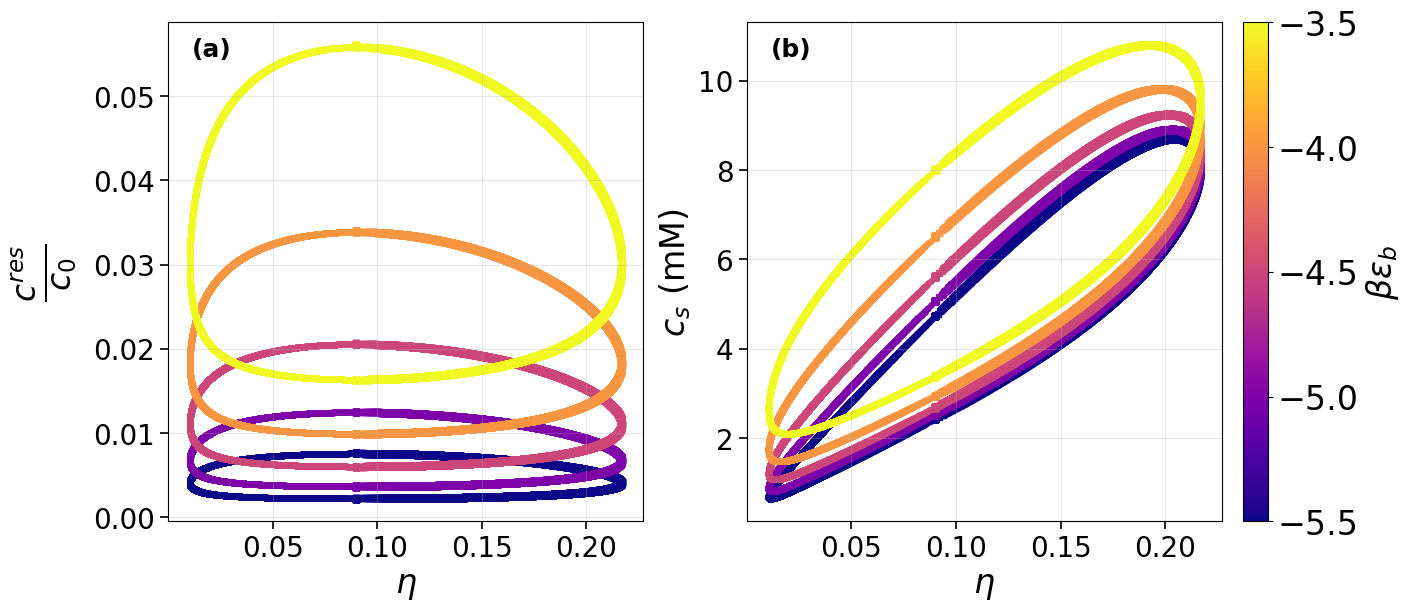}
    \end{center}
    \caption{Change of binodal loop in the single-type-of-patch model upon variation of $\beta \beps_b$ (``squeeze/stretch behavior''). $\beta\epsilon_{uo}=-14$, $c_0$ = 1M, $R_p =2.8$ nm. (a) $\eta$-$\cres$ plane (reservoir salt concentration). (b) $\eta$-$c_s$ plane (physical salt concentration).} 
    \label{fig:shift}
\end{figure}

In Fig.~\ref{fig:shift} the change in the binodal under a variation of the ion-binding energy $\beps_b$ is studied. Under such a change, the maximal coexistence gap $\Delta \eta_\text{max}$ and the critical occupations $\theta_{c,i}$ do not change (since $\eps_\text{pp}$ remains constant). The corresponding critical reservoir salt concentrations are given by
\be
  \cres_{c,i} = c_0\, \frac{\theta_{c,i}}{1-\theta_{c,i}} \, \exp(\beta \beps_b) 
\ee
and are seen to be scaled by $\exp(\beta \beps_b)$. Thus the coexistence loop will become squeezed in the vertical (salt concentration) direction when making $\beps_b$ less negative or stretched in the other case.

\subsection{Two types of patches}
\label{sec:two}

With two types of patches in solutions with a single type of ion, we define two distinct occupation probabilities, $\theta^i_\alpha=\{\theta_1,\theta_2\}$ and two distinct ion-patch binding energies $\beps_b^{i,\alpha}=\{\beps^1_b,\beps^2_b\}$ , with
\be
  \theta_\alpha = \frac{1}{1+\exp\big(\beta({\beps_b^\alpha -\mu_{s})\big)}} \quad (\alpha=1,2) \;.
\ee
This is a parametric equation for a curve in the $\theta_1$-$\theta_2$ plane, elimination of the parameter $\mu_s$ gives
\begin{equation}
    \theta_{2} = \frac{\theta_{1}}{\theta_{1}+(1-\theta_{1})\, \exp(-\beta\,\Delta \beps_b^{1 2})}
    \label{eq:t2_t1} 
\end{equation}
where $\Delta \beps_b^{1 2}=\beps_b^2-\beps_b^1$. 

\begin{figure}
    \centering
    \includegraphics[width=12cm]{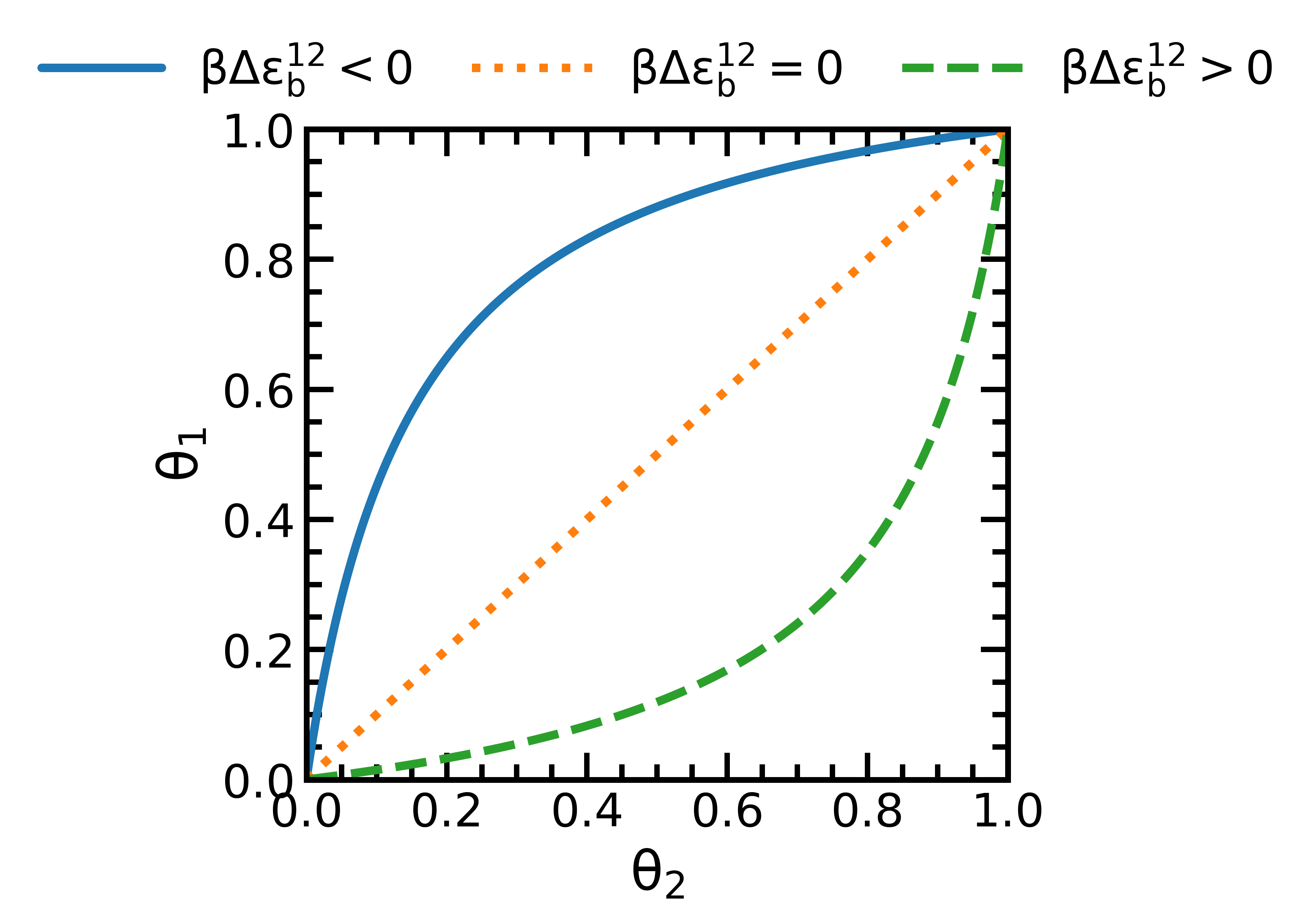}
    \caption{Upon variation of the salt chemical potential, the occupation probabilities $\theta_\alpha$ trace a curve in the $\theta_1$-$\theta_2$ plane, here shown for ion binding energies being equal, $\beps_b^1=\beps_b^2$, and different with $\beta \Delta \beps_b^{1 2}= \pm 2$. 
    }
    \label{fig:t2_t1}
\end{figure}

\begin{figure}[ht!]
  \begin{center}
    \includegraphics[width=18cm]{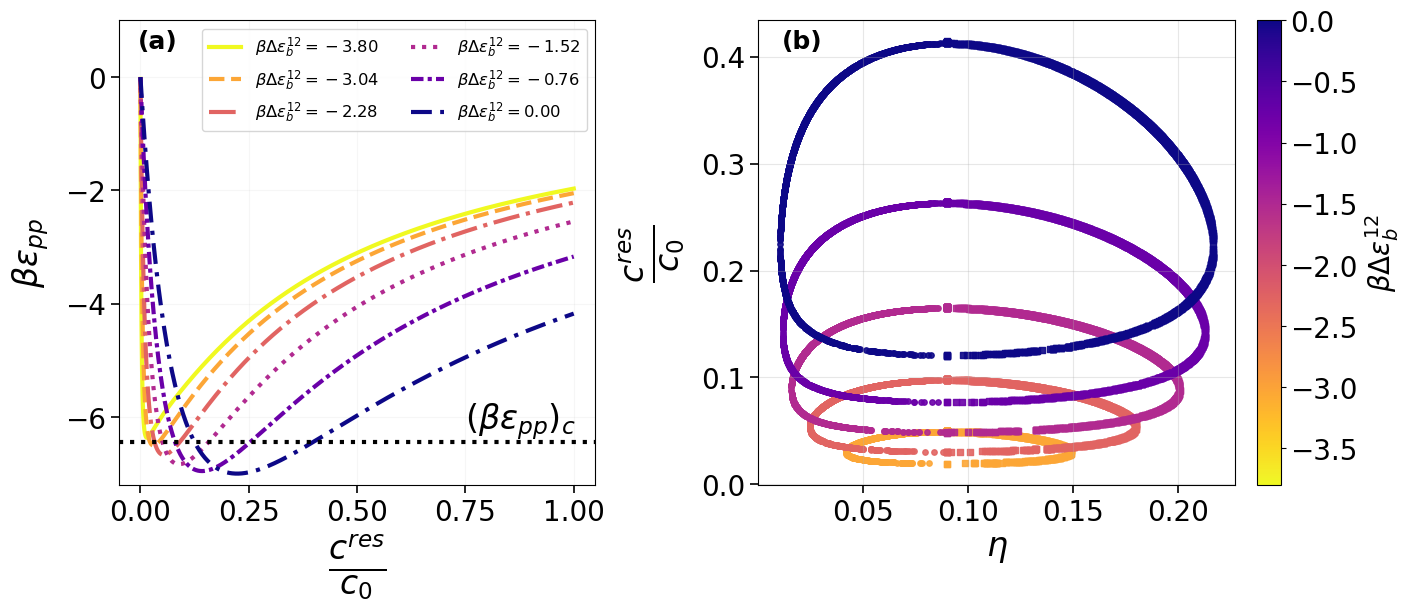}
  \end{center}
    \caption{The case of like-patch attraction dominance, interaction energy parameters are $\beta \eps^{11} = -22$, $\beta \eps^{22} = -12$, $\beta \eps^{12} = -11$. The number of patches is $M=4$, and $m_1=m_2=2$. The difference in ion binding energy $\Delta \beps_b^{1 2}=\beps_b^2-\beps_b^1$ is varied ($\beta \eps^{2}_{b}= -1.5$). 
    (a) Effective patch energy $\beta \eps_\text{pp}(\cres)$ as function of reservoir salt concentration $\cres$. The dashed line is the critical strength for LLPS. 
    (b) Binodal loops in the $\eta$-$\cres$ plane.
    }
    \label{fig:aa_dominance}
\end{figure}

The effective patch-patch energy becomes
\bea
 \label{eq:epp2}
 \eps_\text{pp} &=& \sum_{\alpha=1}^2 a_\alpha\, \theta_\alpha (1- \theta_\alpha)  + b\, (\theta_1+\theta_2-2\theta_1\theta_2) \;, \\
 & & a_\alpha =   2\eps_\text{uo}^{\alpha\alpha}\, \frac{m_\alpha^2}{M^2} \;, \quad   b = 2 \eps_\text{uo}^{12}\, \frac{m_1 m_2}{M^2} \;.\nonumber
\eea
One can distinguish two cases which differ in the conditions when $\eps_\text{pp}$ is most attractive. Assuming $|a_1| \ge |a_2|$:
\bea
   b^2 < a_1 a_2   & \to & \theta_\alpha^\text{min} = \left\{ \frac{1}{2}, \; \frac{1}{2} \right \} \\
   b^2 > a_1 a_2   & \to &  \theta_\alpha^\text{min} = \left \{ \text{min}\left(\frac{1}{2}+\frac{b}{2a_1},\; 1\right) ,0 \right \}   \nonumber \;.
\eea
For $|a_2| > |a_1|$, $\theta_1^\text{min}$ and $\theta_2^\text{min}$ are exchanged and $a_1$ is replaced by $a_2$ in the formula above. 
The first case corresponds to a dominance of attraction between patches of the same type, here the attraction is maximized if the patches have equal occupation probability of 1/2. In contrast, the second case corresponds to a dominance of attraction between patches of different type and here the attraction is maximized if one kind of patch is fully (or nearly fully) occupied by ions and the other kind of patches is empty (or nearly empty). Using the function $\theta_2(\theta_1)$ (see Eq.~(\ref{eq:t2_t1})), one sees under which conditions the attraction can be maximized, see also Fig.~\ref{fig:t2_t1} for characteristic examples of $\theta_2(\theta_1)$. In the first case, the ion binding energies with the two patch types need to be equal which entails $\theta_1=\theta_2$ and thus the maximal attraction at $\theta_1=\theta_2=1/2$ can be realized. In the second case, the ion binding energies need to maximally different, such that one type of patch can be fully occupied and the second patch type empty, thus realizing maximal attraction between unequal patches.       

\begin{figure}[ht!]
\begin{center}
    \includegraphics[width=18cm]{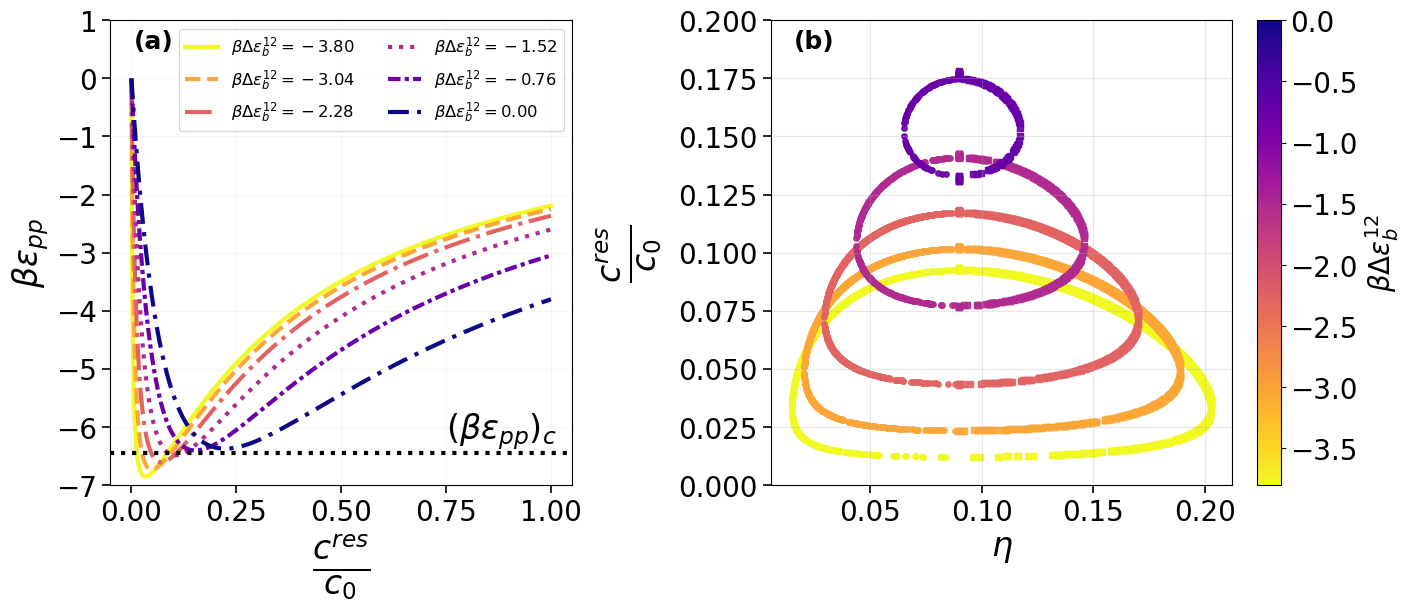}    
\end{center}
\caption{
The case of unlike-patch attraction dominance, interaction energy parameters are $\beta \eps^{11} = -11$, $\beta \eps^{22} = -11$, $\beta \eps^{12} = -14.45$. The number of patches is $M=4$, and $m_1=m_2=2$. The difference in ion binding energy $\Delta \beps_b^{1 2}=\beps_b^2-\beps_b^1$ is varied ($\beta \eps^{2}_{b}= -1.5$). 
    (a) Effective patch energy $\beta \eps_\text{pp}(\cres)$ as function of reservoir salt concentration $\cres$. The dashed line is the critical strength for LLPS. 
    (b) Binodal loops in the $\eta$-$\cres$ plane.
}
    \label{fig:ab_dominance}
\end{figure}

The different behavior of the effective patch-patch attraction to the difference in ion binding energy for the two cases leads also to qualitatively different behavior of the binodals. For case 1 (like-patch dominance), Fig.~\ref{fig:aa_dominance}(a) shows  $\beta \eps_\text{pp}(\cres)$ and Fig.~\ref{fig:aa_dominance}(b) the phase diagram in the $\eta$-$\cres$ plane, all for fixed values of the patch-patch interactions $\eps^{\alpha\beta}$ but different $\Delta \beps_b^{1 2}=\beps_b^2-\beps_b^1$. 
The dashed line in Fig.~\ref{fig:aa_dominance}(a) denotes the critical attraction strength $\tc$, phase separation only occurs if $\beta \eps_\text{pp}(\cres)$ is below the dashed line. For equal binding energies ($\Delta \beps_b^{1 2}=0$) the minimum of $\beta \eps_\text{pp}$ is well below $\tc$ thus defining a large coexistence region. For increasing $\Delta \beps_b^{1 2}$ the minimum moves up and to smaller $\cres$, eventually not crossing the dashed line anymore. For the phase diagram in Fig.~\ref{fig:aa_dominance}(b) this implies that for increasing $\Delta \beps_b^{1 2}$ the binodal loop becomes squeezed in both $\eta$, $\cres$, moves to smaller reservoir salt concentrations and finally disappears.

In case 2 (attraction between unlike patches dominates), the behavior is different, see Fig.~\ref{fig:ab_dominance}. The effective energy $\beta \eps_\text{pp}(\cres)$ has the deepest minimum for the largest value of $\Delta \beps_b^{1 2}$ (Fig.~\ref{fig:ab_dominance}(a)) since the effective attraction is strongest if one kind of patch is occupied and the other one empty. The corresponding coexistence region is large and located at small salt reservoir concentrations (Fig.~\ref{fig:ab_dominance}(a)). Lowering $\Delta \beps_b^{1 2}$, the difference between ion binding energies, the minimum in the effective energy moves up towards $\tc$ and disappears for $\Delta \beps_b^{1 2}=0$. The corresponding binodal loops move up in $\cres$ and become noticeably squeezed in the $\eta$-variable until they disappear.     

In contrast to the changes in phase behavior introduced by varying the ion-binding energies, the opposite case of varying the patch-patch energies with ion-binding energies held constant does not introduce new effects. If the overall effective patch-patch energy is changed (by scaling all $\eps^{\alpha\beta}$ in the same way), the binodals change in the onion-shell manner seen in Fig.~\ref{fig:onion}. {But also if one varies e.g. only $\eps^{12}$ to induce a crossing from like-patch dominance to unlike-patch dominance, the onion-shell behavior is preserved.  }

\begin{figure}[ht!]
    \centering
    \includegraphics[width=0.99\linewidth]{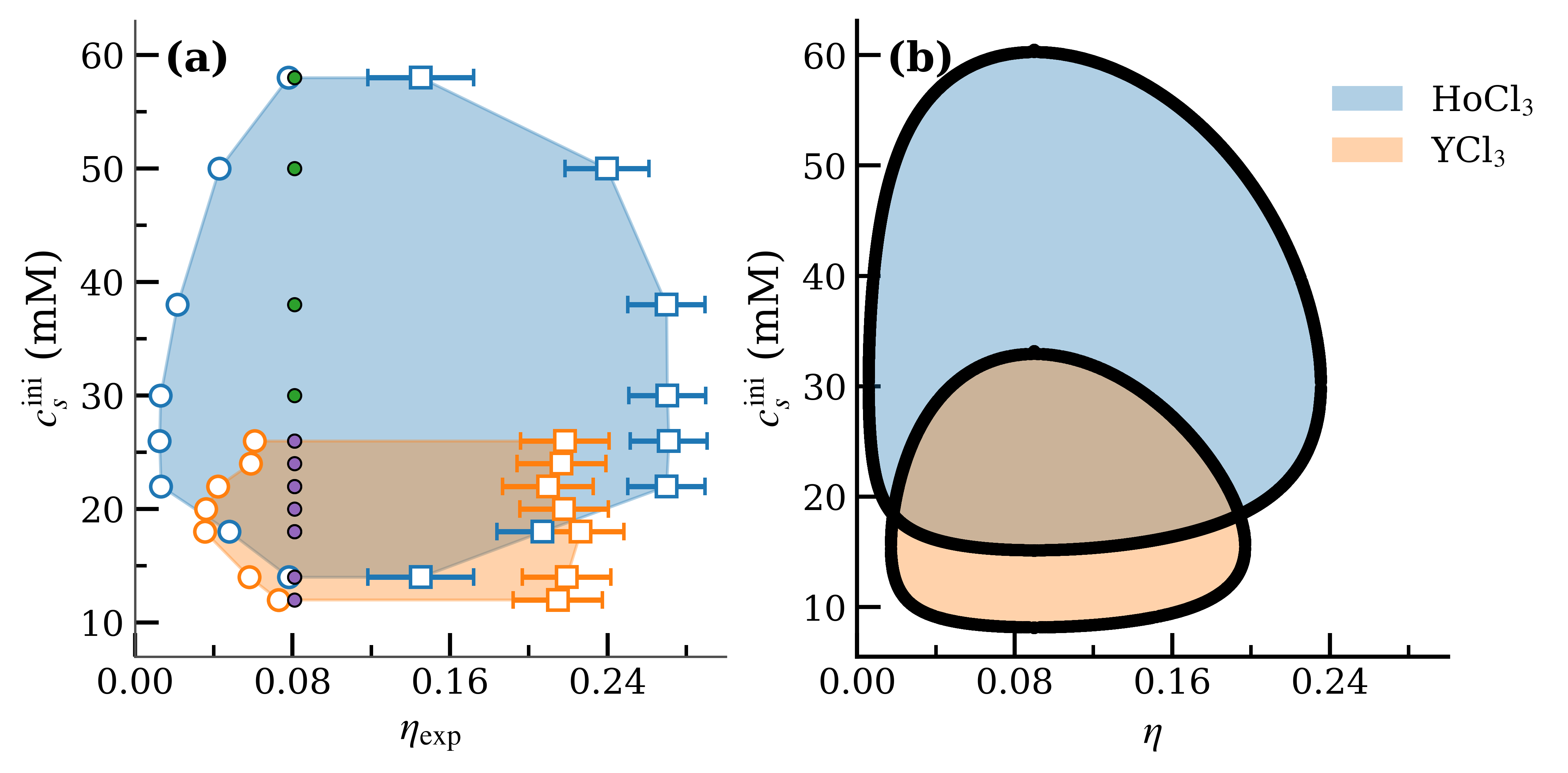}
    \caption{(a) Experimental phase diagram in the $\eta_\mathrm{exp}$-$c_s^\text{ini}$ plane for BSA in the presence of HoCl$_3$ (light blue) and YCl$_3$ (orange), open circles denote the coexisting dilute phase, open squares the coexisting dense phase. Filled points represent sample preparation conditions. (b) Theoretical phase diagram dominated by $\beta\eps_{11}$ (like patch) interactions. Parameters are $\beta\eps_{11}=-25$, $\beta\eps_{22}=\beta\eps_{12}=-11$, with a total number of patches $M=4$. 
    Ion binding energies for the HoCl$_3$ case (blue) are $\beta\eps^1_b=\beta\eps^2_b=-3.5$, while in the YCl$_3$ case (orange)  they are $\beta\eps^1_b=-5.1$ and $\beta\eps^2_b= -2.3$ with $m_1=2$ and $m_2=2$.
    }
    \label{fig:phasediag_exp_theo}
\end{figure}

\subsection{Salt-dependent demixing from experiment}
\label{sec:exploop}

Using the experimental procedure described in Sec.~\ref{sec:exp} phase diagrams in the $\eta_\mathrm{exp}$-$c_s^\text{ini}$ plane have been determined where $\eta_\mathrm{exp}$ is the experimental protein packing fraction (determined from its specific volume and corresponding to a  packing fraction of hard spheres with radius 2.7 nm)  and 
$c_s^\text{ini}$ is the initial salt concentration (the salt concentration in the coexisting phases could not be determined). The two phase diagrams (for HoCl$_3$ and YCl$_3$, respectively) are shown in Fig.~\ref{fig:phasediag_exp_theo}. They are similar in shape to the theoretical phase diagrams in the $\eta$-$\cres$ plane, see e.g. Figs.~\ref{fig:shift}(a) and \ref{fig:aa_dominance}(b). Compared to the broad loop in the HoCl$_3$ case, the binodal for the BSA-YCl$_3$ system is shifted downwards to lower salt concentrations and shrunk in both salt concentration and protein packing fraction directions. This suggests a theoretical description where the main difference between the Ho$^{3+}$ and Y$^{3+}$ ions lies in their binding energies to the patches on the BSA surface. We have evaluated such phase diagrams (for two types of patches) in the $\eta$-$c_s^\text{ini}$ plane where the initial salt concentration follows from Eq.~(\ref{eq:cs}):
\begin{equation}
   c_s^\text{ini}  =  (m_1 \theta_1+m_2\theta_2) \rho^\text{ini} + \cres(\mu^i_s) (1 - \eta^\text{ini} (1 + R_s/R)^3) \;,
\end{equation}
where $R/R_s=18$, $\eta^\text{ini}=0.08$ ($\rho^\text{ini}=\eta^\text{ini}/(4\pi R^3/3)$) is the experimental initial protein volume fraction, and $\theta_{1[2]}$ are the two patch occupation probabilities for coexisting points.
Indeed, the two binodal loops can be fitted almost quantitatively with the model having $M=4$ patches (with $m_1=m_2=2$),
patch-patch binding energies are $\beta\eps_{11}=-25$, $\beta\eps_{22}=\beta\eps_{12}=-11$.
The ion-patch binding energies are different for the two types of patches in the case of YCl$_3$ ($\beta\eps^1_b=-5.1$ and $\beta\eps^2_b= -2.3$), for HoCl$_3$ the binding energy is $\beta\eps_b=-3.5$ for all patches.
This is consistent with a picture of a more polarizable ion such as Ho$^{3+}$ (less specific) and a more compact hard ion such as Y$^{3+}$ (more specific).

Our findings align with previous observations on the effects of various salts in BSA--Me$^{3+}$ systems\cite{matsarskaia2016cation,Matsarskaia_2018_PhysChemChemPhys}.
There, the propensity to phase separation was measured by evaluating the reduced second virial coefficient $B_{2}/B_{2}^{\mathrm{HS}}$ as a function of initial salt concentration for a given initial protein concentration. Phase separation was linked to $B_{2}/B_{2}^{\mathrm{HS}}<-1.5$ (Noro-Frenkel criterion) and it turned out that
Ho$^{3+}$ showed the largest salt concentration interval fulfilling this criterion, Y$^{3+}$ shows a smaller interval, and La$^{3+}$ is the weakest salt (inducing attractions but with $B_{2}/B_{2}^{\mathrm{HS}}>-1.5$), i.e. the system does not undergo LLPS under the tested conditions\cite{Matsarskaia_2018_PhysChemChemPhys}.
For a thorough discussion on applying this criterion to BSA solutions, see Ref.~\cite{weimar2024effective}.

\section{Summary and conclusion}
\label{sec:summary}

In this work, we have extended the ion-activated patchy particle model of Ref.~\cite{roosen2014ion} to account for multiple patch types which reflects the heterogeneity of a protein surface and thus also the varying strength of protein surface patches to bind ions. Generically, the model gives rise to liquid-liquid phase separation described by closed binodal loops in the plane spanned by the protein and salt concentrations. Already for the case of two patch types, the variation of the ion-binding energy leads to characteristic effects on the binodal loop where one may distinguish between two cases. If the attraction between occupied and unoccupied patches is dominated by pairs of like patches, an increasing difference of ion-binding energies between the two patch types leads to a shrinking and eventually a disappearance of the binodal loop towards lower salt concentrations. In the other case of dominant attraction between unlike patches, the binodal loop has a large size if the ion-binding energies are very different and it shrinks and moves to higher salt concentrations if the ion-binding energies get closer in magnitude. The underlying mechanism in the model for this behavior is that by varying the salt chemical potential, a line in the plane of the occupation probabilities for ions on patches is described which may visit different spots in a landscape for an effective patch-patch attraction energy. Upon an increase in ion-binding energy difference between the patch types the line moves away from the attractive minimum (like-patch dominance) or closer to it (unlike-patch dominance). In comparison
with models with non-activated patches, where the gas-liquid transition disappears when
the number of patches approaches two \cite{bianchi2006phase}, we find the complementary mechanism that ions
may shift the attractions from stronger to weaker patches and vice versa.                     

Experimentally, a phase diagram has been determined for a solution of BSA proteins with HoCl$_3$ and YCl$_3$. In comparison with the HoCl$_3$ case, the binodal loop for the YCl$_3$ case is shrunk and shifted towards lower salt concentration. This corresponds to the case of like-patch dominance and both experimental phase diagrams could be described very well by the model using reasonable parameters.    

In conclusion, we have obtained novel insight into possible mechanisms of LLPS in protein solutions with trivalent salts which arise by ion-specific effects, as seen in experiment and expressed in the present model by different binding energies of the ions (depending on the specific ion) towards patches on the protein surface. Although the model has simplified many atomistic details of a protein solution with trivalent salt ions, we believe it provides important clues which warrant further consideration. In the future, it would be desirable to link the model parameters to more explicit microscopic investigations such as atomistic simulations, an example e.g. is Ref.~\cite{sahoo2022simulationbsa} which investigates cation-protein binding free energy through simulations.        

\section*{Acknowledgments}
The authors acknowledge the Deutsche Forschungsgemeinschaft (DFG, PATMI2.0, SCHR700/48-1) 
for funding, which made this research possible. We also thank Prof. Dr. Roland Roth and Melih Gül for fruitful discussions.

\section*{Statements}

\textbf{Conflict of Interests:} The authors have no conflicts to disclose.

\textbf{Data Availability:} The data that support the findings of this study are available from the corresponding author upon reasonable request.

\textbf{Author contributions:} Conceptualization - Frank Schreiber, Martin Oettel; Formal Analysis - Furio Surfaro, Martin Oettel; Funding Acquisition – Frank Schreiber; Investigation - Furio Surfaro, Peixuan Liang, Hadra Banks, Martin Oettel; Methodology - Fajun Zhang, Frank Schreiber, Martin Oettel; Writing/Original Draft Preparation - Furio Surfaro, Martin Oettel; Writing/Review and Editing -  Furio Surfaro, Fajun Zhang, Frank Schreiber, Martin Oettel

\begin{appendix}
 \section{Wertheim theory for the effective model}
 \label{app:wertheim}

\begin{figure}
    \centering
   \includegraphics[width=8cm]{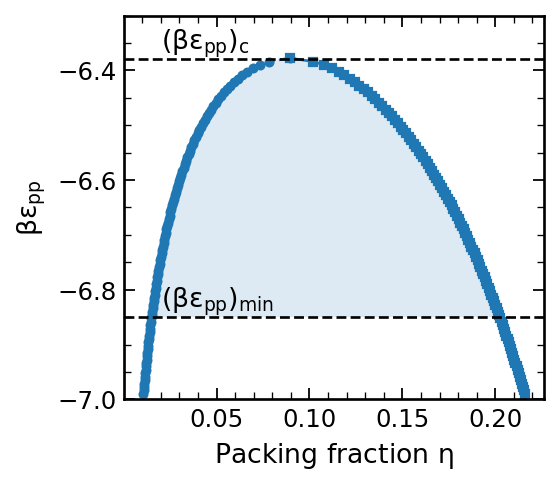}
    \caption{Binodal from Wertheim theory, using the following parameters: $M=4$, $K=0.002376\,R^3$. }
    \label{fig:binodal}
\end{figure}

In Wertheim theory \cite{Wertheim1984a,Wertheim1984b,Wertheim1986}, the free energy density is the sum of the free energy density of the reference system and a perturbative contribution from particle bonding. The hard-sphere fluid with the free ergy density from the Carnahan-Starling equation of state \cite{carnahan1969} is used as the reference system. 
The bonding contribution to the free energy per volume 
is denoted as $f_\text{bond}$:
\begin{equation}
    \beta f_\text{bond} = M \, \frac{\eta}{\nu_s} \left(\ln(X)-\frac{X}{2} +\frac{1}{2}\right),
\end{equation}
Here, the packing fraction $\eta = 4/3 \pi R^3 \rho = \nu_s \rho$, where $\nu_s$ represents the volume occupied by a single particle, $M$ is the number of patches per particle, and $X$ is the probability that a patch has not formed a bond. Note that $X$ depends on the number density $\rho$, or equivalently $\eta$, and is determined by the mass-action equation \cite{roosen2014ion}:
\begin{equation}
\frac{1-X}{X^2} = m \frac{\eta}{\nu_s} \Delta,
\label{masslaw}
\end{equation}
where $\Delta$ accounts for the spherically averaged interaction between bonded patches of two particles. Following Refs.~\cite{jackson1988,roosen2014ion}, we employ a square-well potential between patches of depth $\eps_\text{pp}$ with a short range. After performing the angular average, $\Delta$ can be written as
\begin{equation} \label{eq:delta}
\Delta = 4 \pi g_{HS}(\sigma, \eta) K F,
\end{equation}
where $g_{HS}(\sigma, \eta)$ is the contact value of the radial distribution function for the hard-sphere reference system with diameter $\sigma$, $K$ is the bonding volume, and 
$F=\exp(-\beta \epsilon_\text{pp}) - 1$ comes from the angular average of the Mayer-$f$ function for the patch-patch interaction. In Ref.~\cite{roosen2014ion}, $K=0.002376\,R^3$ was employed which is also the value we use here.  

The binodal is obtained from the equality of chemical potential and pressure for the coexisting states and is shown in Fig.~\ref{fig:binodal}. The dimensional critical interaction strength is $\tc \approx - 6.38$. In the main text, the dimensionless occupation-dependent effective interaction strength $\beta \eps_\text{pp}$ is defined in Eq.~(\ref{eq:beta_epp_general_symmetric}) and it has a minimum  $(\beta \eps_\text{pp})_\text{min}$ along the trajectory defined by the occupation probabilities $\theta_i^\alpha$. If $(\beta \eps_\text{pp})_\text{min}< \tc$, phase coexistence as defined by the blue region occurs (see Fig.~\ref{fig:binodal}). In Ref.~\cite{roosen2014ion}, a single type of patch was considered with $\beta \eps_\text{uo}=-14$ which leads to  $(\beta \eps_\text{pp})_\text{min}=-7$, and the resulting width of the phase coexistence loop in the $\eta$ variable was shown to be in reasonable agreement with experimental results for a HSA-YCl$_3$ system.

\section{Wertheim theory for an explicit model with occupied and unoccupied patches treated as two different binding sites}
\label{app:twobindingsites}

As explained in the beginning of Sec.~\ref{sec:theory}, we treated the particle-linker system of proteins and ions with two approximations: (i) the statistics of protein-ion association was treated using a two-level system and (ii) an effective patch-patch energy was introduced (Eq.~(\ref{eq:beta_epp_general_symmetric})) which averaged over all pairwise patch-patch interactions (occupied and unoccupied). With regard to approximation (ii), one might be worried that this average (where the energies enter linearly) is not adequate within Wertheim theory, since the attractive patch-patch energies enter exponentially (see Eq.~(\ref{eq:delta}) and text below). We check this by investigating a variant of the model (extended ion-activated patchy particle) in which particles possess $\mo=\theta M$ occupied and $\muo=(1-\theta)M$ unoccupied patches and only an attractive interaction between those are considered (here not depending on any patch type). The number of occupied and unoccupied patches ($\mo$ and $\muo$) might be non-integer but this poses no problem in applying Wertheim theory for two patch types \cite{jackson1988,fantoni2015wertheim}. The model has the following mass law equations:      


\begin{align}
\xo &= \frac{1}{1 + \muo \, \rho \, \xu \, \Delta}\, \\
\xu &= \frac{1}{1 + \mo \, \rho \, \xo \, \Delta} \;.
\end{align}
Here, $\xo$ and $\xu$ are the fractions of nonbonded occupied and unoccupied patches. 
%
\begin{eqnarray}
\xu &=& \frac{\rho\, \Delta (\muo - \mo) -1 + \sqrt{\left(1 + \rho\, \Delta (\mo - \muo) \right)^2 + 4\,\muo\, \rho\, \Delta} }{ 2 \muo\, \rho\, \Delta}
\label{eq:a} \\
\xo & =&  1 - \frac{\muo}{\mo} + \frac{\muo}{\mo} \xu \;.
\label{eq:b}
\end{eqnarray}
The resulting bonding free energy 
contains additively the contributions from the bonded and the unbonded sites:
\begin{equation}
 \beta \, f_\text{bond} = \rho \, \mo\, \left(\ln \,\xo- \frac{\xo}{2} + \frac{1}{2} \right) + \rho \, \muo\, \left(\ln \,\xu- \frac{\xu}{2} + \frac{1}{2}\right)
    \label{eq:free_extended}
\end{equation}
Note that completely equivalent expressions have been obtained in discussing the association problem of a nanoparticle with $N_L$ ligands that can bind to $N_R$ receptors on a surface such that only one ligand may be bound to one receptor with a free energy weight $\chi=\exp(-\beta G)$ (where $G$ is a bond Gibbs energy ) \cite{angioletti2020}. With the identification $N_L \equiv \rho\muo$, $N_R \equiv \rho\mo$ and $\chi \equiv\Delta$ the corresponding equations (13) and (14) in Ref.~\cite{angioletti2020} are found.

\begin{figure}[ht!]
    \centering
    \includegraphics[width=10cm]{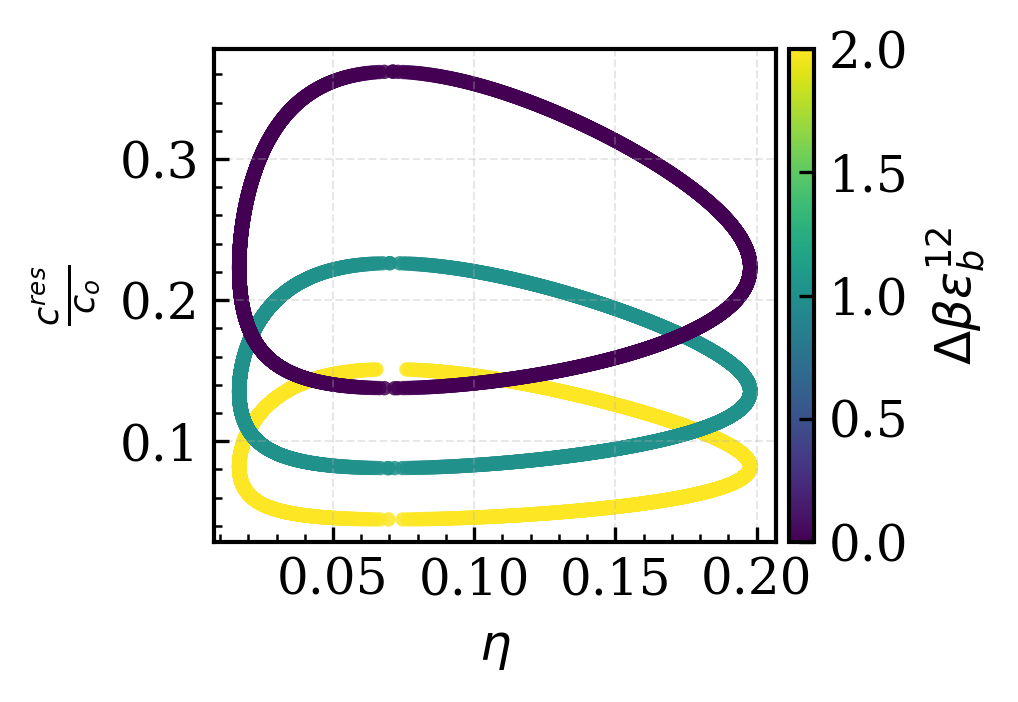}
    \caption{Effect of changing the relative binding energy $\Delta\beta\epsilon^{12}_b$ for a system with $M=m_1+m_2=4$ with $m_1=m_2=2$. Phase coexistence loops change similarly to the "squeeze/stretch behavior" in Fig.~\ref{fig:shift} for a single patch type.}
    \label{fig:extended}
\end{figure}

As an example, we study numerically the case that two patch types (with numbers $m_1$ and $m_2$) exist with regard to the ion-binding affinity, expressed by two distinct occupation probabilities, $\theta_\alpha=\{\theta_1,\theta_2\}$ and two distinct ion-patch binding energies $\beps_b^{\alpha}=\{\beps^1_b,\beps^2_b\}$. However, the patch-patch binding energy between occupied and unoccupied patches does not distinguish between different $\alpha$, i.e. $\eps^{\alpha\beta}_\text{uo}=\eps$. In the above expressions, we need to modify only $\mo \to \sum_\alpha m_\alpha \theta_\alpha $ and $\mo \to \sum_\alpha m_\alpha (1-\theta_\alpha)$.     
In Fig.~\ref{fig:extended}, we show phase diagrams in the $\eta$-$\cres$ plane for $\beta \eps=-7.5$, $\beta \beps_b^2= -1.5$ for different values of $\Delta \beps_b^{1 2}=\beps_b^2-\beps_b^1$. The behavior with increasing $\Delta \beps_b^{1 2}$ is very similar to the ``squeeze/stretch behavior'' of the model with single type of patch (see Fig.~\ref{fig:shift}). With respect to the patch-patch attractive energy, it is of single type, and therefore the disappearance of the phase coexistence loop associated with either the dominance of like patch attractions or of unlike patch attractions is not seen. We note that in the extended model a patch-patch attractive energy only slightly below $\tc$ is needed, and that the coexistence loop is more asymmetric with a shifted location of the critical points (as compared to the one-patch version of the ion activated model of the main text. Otherwise the qualitative behavior agrees with it.



\end{appendix}

\bibliography{References}
\end{document}